\documentclass[
preprint,
tightenlines
,showpacs
,showkeys
,nofootinbib
,aps
,amsfonts
,amssymb
,pdftex  
]{revtex4-1}

\usepackage{graphicx}  
\usepackage{amsmath, amssymb, bm}   
\usepackage[dvips]{color}
\usepackage[normalem]{ulem}  
\usepackage{pslatex}  
\usepackage{psfrag,epsfig}
\usepackage{subfigure}

\begin{document}

\title{
Short-distance RG-analysis of $X(3872)$ radiative decays}

\author{%
D. A. S. Molnar}
\email{dmolnar@if.usp.br}
\affiliation{Instituto de F\'\i sica, Universidade de S\~ao Paulo,
C.P. 66318, 05314-970, S\~ao Paulo, SP, Brazil}

\author{%
R. F. Luiz}
\email{rafael.fernandes.luiz@usp.br}
\affiliation{Instituto de F\'\i sica, Universidade de S\~ao Paulo,
C.P. 66318, 05314-970, S\~ao Paulo, SP, Brazil}

\author{%
R. Higa}
\email{higa@if.usp.br}
\affiliation{Instituto de F\'\i sica, Universidade de S\~ao Paulo,
C.P. 66318, 05314-970, S\~ao Paulo, SP, Brazil}

\begin{abstract}
We present a renormalization-group analysis of the $X(3872)$ radiative 
decays into $J/ \psi \gamma$ and $\psi(2S) \gamma$. 
We assume a $D\bar{D}^{*}$ molecule for the $X$ long-distance structure and 
parametrize its short-distance physics as contact interactions. 
Using effective field theory techniques and power-divergence subtraction 
scheme, we find that short- and long-distance physics are equally important 
in these decays. Our calculations set a lower limit to the corresponding 
decay widths, which can in principle be tested experimentally. 
Our results may be used as guide to build models for the $X$ short-distance. 
\end{abstract}

 
\date{\today}

\maketitle

\section{Introduction}

The interest in the physics of heavy quarkonia was spurred, more than a 
decade ago, by the discovery of the exotic meson 
$X(3872)$~\cite{belle,cdf,d0}, 
whose physical properties do not fit in the historically well-succeeded 
quark model. 
Soon after, a new ``particle zoo'' of exotic particles was uncovered by 
several collaborations around the world. 
The challenge imposed by the so-called $X$, $Y$, and $Z$ states triggered 
alternative explanations for their structures, such as tetraquarks, 
molecules, gluonic excitations, mixtures, and 
others~\cite{Bauer2005, Swanson2006, Nielsen2009, Brambilla2010}. 
The latest discoveries of charged mesons in the charmonium and bottomonium 
sectors put in check explanations based on either the conventional quark 
model, or hybrids of charmonium and gluonic excitations since, besides the 
$c\bar c$ pair, two additional quarks must exist to provide the charge of 
these states. 
These recent findings highlight the complexities of the strong force in its 
non-perturbative regime, 
with the promise of improve our understanding about 
the emergence of confinement in QCD. 

Despite the large amount of theoretical investigations in the literature, 
very little is known about the $X(3872)$ structure 
(denoted by $X$ from now on). 
However, the purely molecular interpretation is very appealing, since it 
has a mass remarkably close to the $D \bar{D}^{*}$ 
threshold~\cite{Voloshin:1976ap,DeRujula:1976zlg,Tornqvist:2004qy} and 
its small width is not easily accommodated in the conventional quarkonium 
picture. 
On the experimental side, only recently the $X$ quantum numbers were 
confirmed as $J^{PC}=1^{++}$ by the LHCb collaboration~\cite{cp}. This result 
practically rules out any conventional charmonium explanation, albeit 
not enough to distinguish among the remaining exotic possibilities. 

Swanson, in Ref.~\cite{swanson}, suggested looking at the $X$ 
radiative decays into $\gamma J/ \psi$ and $\gamma \psi(2S)$ as one of the 
promising tests for its molecular nature. 
His molecular model calculation includes both neutral and charged 
$D\bar D^{\star}$ states, decaying to a photon via light quark 
annihilation, and smaller components of $\rho J/\psi$ and $\omega J/\psi$ 
decaying to a photon via vector-meson dominance. He obtains 
$\Gamma(X\to\gamma J/\psi)=8$~keV and $\Gamma(X\to\gamma \psi')=0.03$~keV. 
His quark model calculations, though very sensitive to the detailed 
assumptions involved, give $\Gamma(X\to\gamma J/\psi)\sim 70$-$140$~keV and 
$\Gamma(X\to\gamma \psi')\sim 95$~keV. Based on these results, the author 
claims that radiative decays, especially decaying into the $\psi(2S)$ 
channel (here denoted by $\psi'$), would be a sharp test for the $X$ 
molecular nature. In 2009 BaBar measured the ratio
\begin{equation}
R \equiv \frac{\Gamma[X \to \gamma \psi']}
{\Gamma[X \to \gamma J/ \psi]}
\label{eq:ratio}
\end{equation}
obtaining $R = 3.4 \pm 1.4$ \cite{babar}. 
Later attempt to measure the same ratio was done by the Belle collaboration. 
They could not find any signal though, setting only the upper limit 
$R < 2.1$ at $90 \%$ confidence level \cite{belle2}. 
Recently, the LHCb collaboration reported $ R = 2.46 \pm 0.64 \pm 0.29$, 
where the first uncertainty is statistical and the second is 
systematic \cite{Aaij:2014ala}. The latter concludes that the experimental 
result does not support the pure molecular picture, favoring 
charmonium~\cite{wang,barnes1,barnes2,badalian1,badalian2} 
or mixtures of molecule and charmonium~\cite{nielsen1,nielsen2,taki}. 
However this conclusion, which is based on the results of 
Ref.~\cite{swanson}, is disputable, as indicated by further calculations 
in the molecular approach. 

Using phenomenological meson Lagrangians and assuming the $X$ to be a 
loosely-bound $D^{0}\bar D^{*0}$ molecule, 
Dong and collaborators~\cite{dong} calculated the radiative decay 
$X\to\gamma J/\psi$, obtaining an upper limit of $118.9~{\rm keV}$, 
compatible with some quark model predictions~\cite{swanson}. 
Their calculation is not very sensitive to variations on the binding energy, 
but depends on their form-factor parameter $\Lambda_M$. 
They conclude that their decay width is fully compatible with a 
predominantly molecular nature of X, allowing a very small admixture of 
$c \bar{c}$. 
In a latter work~\cite{Dong:2009uf}, the same authors addressed, besides a 
couple of hadronic decays, radiative decays into both $J/\psi$ and $\psi'$ 
channels. 
Their values for the ratio $R$ were compared against the experimental one 
available at the time, from BaBar~\cite{babar}. 
With essentially the same molecular approach used before~\cite{dong}, but 
with different quark model approaches for their $c\bar c$ component, 
they find a subtle interplay between these two components, depending on 
the $c\bar c$ model and the $X$ binding energy. 

Guo {\it et al.} \cite{guo} investigated the imprints of the long-distance 
$D\bar D^{\star}$ molecular structure of $X$ on the radiative decays 
$X\to\gamma J/\psi$ ($\Gamma_{\gamma J\psi}$) and $X\to\gamma\psi'$ 
($\Gamma_{\gamma\psi'}$), with an effective field theory approach. 
Contrary to~\cite{swanson}, they conclude that the radiative decays do not 
allow one to draw conclusions about the nature of the $X$. 
However, their analysis focus only at the long-distance loop contributions 
to the radiative amplitude, without explicitly considering the 
short-distance contributions, parametrized as contact-like interactions in 
the effective Lagrangian. The main purpose of this work is to perform 
a proper renormalization-group analysis of both loop and contact 
contributions to these radiative decays, therefore, complementing the 
studies of Ref.~\cite{guo}. 

Nevertheless, we adopt here a different prescription to regularize the loop 
integrals, the power-divergence subtraction (PDS) 
scheme~\cite{Kaplan:1998tg,Kaplan:1998we}. 
Devised to handle the non-perturbative aspect of the nucleon-nucleon 
interaction, it soon became an alternative regularization method in other 
non-perturbative systems such as cold-atoms, exotic mesons, and nuclear 
clusters~\cite{Braaten:2004rn,Braaten:2005jj,Rupak:2011nk,
Fernando:2011ts,Fernando:2015jyd}. 
Based on dimensional regularization, the method consists of subtractions, 
beyond the $D=4$ dimensions of the usual modified minimal subtraction 
($\overline{\rm MS}$) scheme, at 
lower dimensions to account for power-law divergences~\cite{Phillips:1998uy}. 
The latter are required in order to guarantee non-trivial 
renormalization-group properties characteristic of weakly-bound 
systems~\cite{Birse:1998dk,Phillips:1997xu}, like the scaling limit and the 
Efimov effect~\cite{Braaten:2004rn}. In fact, for non-perturbative 
systems the usual $\overline{\rm MS}$ scheme seems to fail in reproducing 
a non-trivial scattering amplitude~\cite{Phillips:1997xu}. 

\section{Radiative Decay Amplitudes} \label{amp}

\begin{figure}[hbt]
\begin{center}
\includegraphics[width=0.80\textwidth]{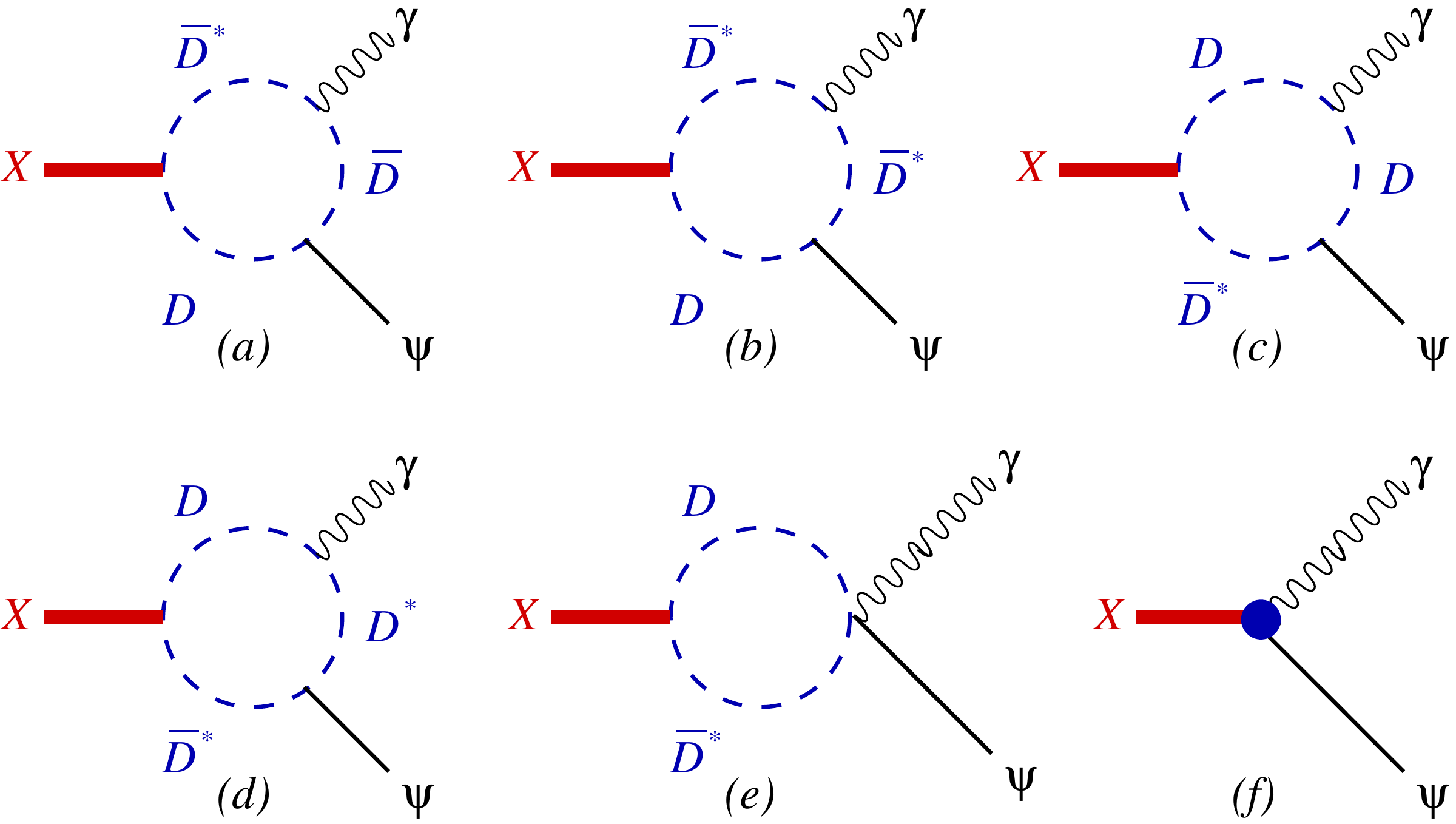}
\caption{Diagrams that contribute to the $X$ radiative decay. 
See text for details.}
\label{fig:diags}
\end{center}
\end{figure}

The interactions are derived from effective Lagrangians based on heavy-meson 
and chiral symmetries~\cite{burdman,alfiky,bira,guo,guo3,Colangelo:2003sa}. 
The loop diagrams $(a)$-$(e)$ from Fig.~\ref{fig:diags} where calculated in 
Ref.~\cite{guo}, and for the sake of completeness we reproduce below the 
relevant expressions. 

One denotes $\psi$ generically for both $J/\psi$ and $\psi'$. The hidden-charm 
mesons $X$ and $\psi$ have masses $M_X$ and $M_{\psi}$ while 
the open-charm mesons $D^{\star}$ and $D$ have masses $m_{\star}$ and $m$, 
respectively. 
The $X$ long-range structure is coupled to a $D\bar D^{\star}$ pair 
with strength $x=x_{\rm nr}\sqrt{M_Xm_{\star}m}$ via the interaction 
Lagrangian 
\begin{equation}
{\cal L}_X=\frac{x}{\sqrt{2}}X^{\dagger}_{\sigma}
\Big(\bar D^{\star\,\sigma}D+\bar DD^{\star\,\sigma}\Big)+\mbox{H.c.}\,,
\label{eq:LagX}
\end{equation}
where the open-charm meson fields stand for both neutral and charged 
ones~\cite{guo}, and 
$x_{\rm nr}^{(0)}=0.97^{+0.40}_{-0.97}\;{\rm GeV}^{-1/2}$ 
was obtained from a non-relativistic relation with the binding 
energy~\cite{guo2}. We define $x_{\rm nr}^{(0)}\equiv x_{\rm nr}(\mu_0)$, 
the renormalized coupling of $X$ to the $D\bar D^{\star}$ pair at the 
scale $\mu_0=M_X$. 

The interaction of the open-charm mesons with the hidden-charm $\psi$ reads 
\begin{eqnarray}
{\cal L}_{\psi}&=&i\,\psi^{\mu\dagger}\bigg\{
g_{\bar DD}\big(\bar D\stackrel{\leftrightarrow}{\partial_{\mu}}D\big)
-g_{\bar D^{\star}D}\,\epsilon_{\mu\nu\alpha\beta}\Big[
\big(\partial^{\alpha}\bar D^{\star\nu}\big)\big(\partial^{\beta}D\big)
-\big(\partial^{\beta}\bar D\big)\big(\partial^{\alpha}D^{\star\nu}\big)
\Big]
\nonumber\\&&
-g_{\bar D^{\star}D^{\star}}\Big[
\bar D^{\star}_{\nu}\stackrel{\leftrightarrow}{\partial_{\mu}}D^{\star\nu}
+\big(\partial_{\nu}\bar D^{\star}_{\mu}\big)D^{\star\nu}
-\bar D^{\star\nu}\big(\partial_{\nu}D^{\star}_{\mu}\big)
\Big]
\bigg\}+\mbox{H.c.}\,,
\label{eq:LagDDpsi}
\end{eqnarray}
with the couplings related to a single parameter $g_2$ via heavy-quark 
symmetry~\cite{guo,guo3,Colangelo:2003sa}: 
\begin{equation}
\frac{g_{\bar DD}}{\sqrt{M_{\psi}}}\frac{1}{m}=
\frac{g_{\bar D^{\star}D}}{\sqrt{M_{\psi}}}\sqrt{\frac{m_{\star}}{4m}}=
\frac{g_{\bar D^{\star}D^{\star}}}{\sqrt{M_{\psi}}}\frac{1}{m_{\star}}
=g_{\psi}\,.
\label{eq:g-HBrels}
\end{equation}

Interactions with the emitted photon have two distinct origins. 
Electric interactions have no additional parameters. They contribute to 
diagrams $(b)$, $(c)$, and $(e)$, via minimal substitution 
$\partial_{\mu}\to\partial_{\mu}+ieA_{\mu}$ 
in the kinetic term of the $D^{\star}$, kinetic term of the $D$, and 
interaction term of the $\bar D^{\star}D\psi$ Lagrangians, respectively. 
Magnetic interactions are derived from the covariant generalization of the 
non-relativistic heavy-meson 
Lagrangian~\cite{guo,Amundson:1992yp,Cheng:1992xi},
\begin{eqnarray}
{\cal L}_{m}&=&ie\,m_{\star}F_{\mu\nu}D^{\star\mu\dagger}_{i}
\Big(\beta Q_{ij}-\frac{Q_c}{m_c}\delta_{ij}\Big)D^{\star\nu}_{j}
\nonumber\\&&
+e\,\sqrt{mm_{\star}}\epsilon_{\lambda\mu\alpha\beta}
v^{\alpha}\partial^{\beta}A^{\lambda}\Big[D^{\star\mu\dagger}_{i}
\Big(\beta Q_{ij}+\frac{Q_c}{m_c}\delta_{ij}\Big)D_{j}+\mbox{H.c.}\Big]\,,
\label{eq:LagDmag}
\end{eqnarray}
where $v^{\mu}$ is the four-velocity of the heavy quark with $v^2=1$, 
$Q=\mbox{diag}(Q_u,Q_d)$ is the charge matrix of the light quarks, $Q_c=2/3$ 
and $m_c$ are the charmed quark charge and mass, respectively. 
They contribute to diagrams $(a)$, $(b)$, and $(d)$. The extra 
parameter $\beta\sim 1/(380\,{\rm MeV})$ takes into account the 
non-perturbative dynamics of the light quark inside the charmed 
meson~\cite{guo}. 

The loop amplitudes from diagrams $(a)$-$(e)$ are written as 
\begin{equation}
{\cal M}^{\rm loop}=\frac{e}{\sqrt{2}}\,x_{\rm nr}g_{\psi}m
\sqrt{M_{X}M_{\psi}}\,\varepsilon^{\mu}_{\psi}(p')\varepsilon^{\sigma}_{X}(p)
\varepsilon^{\lambda}_{\gamma}(q)\int\frac{d^4k}{(2\pi)^4}
S^{\nu\sigma}(k)S(k-p)J_{\mu\nu\lambda}(k)\,,
\label{eq:Mloop}
\end{equation}
where $S^{\nu\sigma}$ ($S$) is the $\bar D^{\star}$ ($D$) 
propagator with momentum $k$ ($k-p$). The polarization 
vectors $\varepsilon_{\psi}^{\mu}$, $\varepsilon_{X}^{\sigma}$, and 
$\varepsilon_{\gamma}^{\lambda}$ stand for $X$, $\psi$, and $\gamma$ with 
external momenta $p$, $p'=p-q$, and $q$, respectively. 
The explicit contributions of each diagram to $J_{\mu\nu\lambda}$ 
is given by Eqs.~(24)-(29) of Ref.~\cite{guo}. 

Diagram $(f)$ is the short-distance $X\gamma\psi$ interaction that 
renormalizes the ultraviolet divergences present in (\ref{eq:Mloop}). 
The corresponding amplitude reads 
\begin{equation}
{\cal M}^{\rm cont}=C_{\psi}\,\varepsilon^{\mu}_{\psi}(p')
\varepsilon^{\sigma}_{X}(p)\varepsilon^{\lambda}_{\gamma}(q)
\,\epsilon_{\mu\sigma\lambda\nu}q^{\nu}\,.
\label{eq:Mcont}
\end{equation}
As pointed out in~\cite{guo}, the necessity of having this term to cancel the 
divergences of the loop diagrams means that the $X$ radiative decays are 
sensitive to both its long- and short-distance structure, making them 
unsuitable as probes exclusively of the former. 
Nevertheless, renormalization-group techniques can be used to estimate the 
strength of the short-range interactions at a limited range of energy scale. 
This is the main goal of this work. In Ref.~\cite{guo}, the strength of 
diagram $(f)$ after renormalization, $C_{\psi}^{(r)}$, was set to zero. 
We take advantage of the most recent experimental value of $R$ by the LHCb 
collaboration to obtain the renormalized values of $C_{\psi}^{(r)}$ in 
both $J/\psi$ and $\psi'$ channels. 

\section{Results and Discussions} \label{results}

The partial width of the $X$ radiative decay, when the initial and final 
polarization states are not measured, is given by 
\begin{equation}
\Gamma = \frac{M_{X}^{2} - M_{\psi}^{2}}{48 \pi M_{X}^{3}} |\mathcal{M}|^2,
\label{eq:GammaX}
\end{equation}
where the total amplitude squared reads 
\begin{eqnarray}
|\mathcal{M}|^2 & = & \sum_{\rm all \; pols.}
\mathcal{M}_{\mu'\sigma'\lambda'} \mathcal{M}^{*}_{\mu\sigma\lambda}
\left(\varepsilon^{\sigma'}_{(X)}(p)\varepsilon^{* \sigma}_{(X)}(p)\right)
\left(\varepsilon^{\mu'}_{(\psi)}(p')\varepsilon^{* \mu}_{(\psi)}(p')\right)
\left(\varepsilon^{\lambda'}_{(\gamma)}(q)\varepsilon^{*\lambda}_{(\gamma)}(q)
\right)
\nonumber\\
& = & \mathcal{M}_{\mu'\sigma'\lambda'}\mathcal{M}^{*}_{\mu\sigma\lambda}
\left(\frac{p^{\sigma'}p^{\sigma}}{M_{X}^{2}}-g^{\sigma'\sigma}\right)
\left(\frac{p^{\prime\mu'}p^{\prime\mu}}{M_{\psi}^{2}} - g^{\mu'\mu}\right)
\left(-g^{\lambda'\lambda}\right)\,.
\label{eq:Msquared}
\end{eqnarray}
We made use of the Mathematica software to deal with contractions of the 
Lorentz indices and algebraic manipulations. The loop integrals in 
Eq.~(\ref{eq:Mloop}) were handled with the usual Feynman parametrizations, 
with the remaining integrations solved numerically with a Gauss-Legendre 
quadrature. 

In order to estimate qualitatively the importance of short-distance physics, 
we first compute the ratio $R$ from Eq.~(\ref{eq:ratio}) 
considering only the long-range loop diagrams $(a)$-$(e)$, which we shall 
denote by $R^{\rm loop}$. 
At this point the analysis is similar to the one from Ref.~\cite{guo}. 
It is evident from Eq.~(\ref{eq:Mloop}) that, in this case, the dependence 
on $x_{\rm nr}$ is cancelled in the ratio $R^{\rm loop}$. However, 
$R^{\rm loop}$ remains very 
sensitive to the ratio $(g_{\psi'}/g_{J/\psi})^2$, which is poorly known. 
From the leptonic decay widths of $J/\psi$ and $\psi'$, 
Ref.~\cite{Dong:2009uf} obtains $g_{\psi'}/g_{J/\psi}\sim 1.67$, which is 
the central value that we adopt. We allow a variation around the natural 
band $1\lesssim g_{\psi'}/g_{J/\psi}\lesssim 2.5$, which should account for 
uncertainties in both phenomenological extraction as well as renormalization 
evolution of this ratio. 

\begin{figure}[hbt]
\centering
 \includegraphics[width=7.5cm]{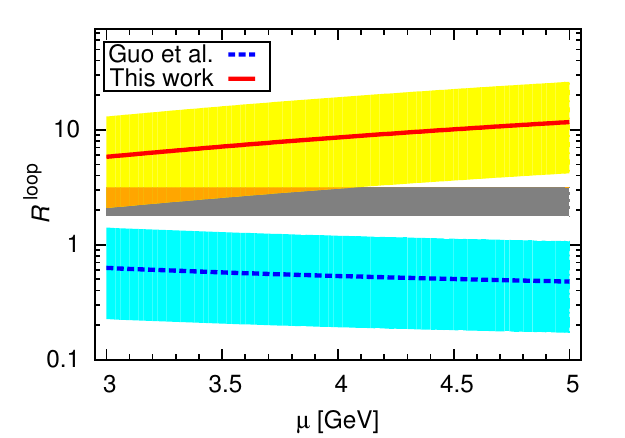}
\caption{Ratio $R^{\rm loop}$ of the branching fractions defined by 
Eq.~(\ref{eq:ratio}). See text for details.}
\label{fig:Ratios}
\end{figure}
Fig.~\ref{fig:Ratios} shows $R^{\rm loop}$ as a function of the 
renormalization scale $\mu$. We choose the interval 
$3\,{\rm GeV}\leq\mu\leq 5\,{\rm GeV}$, not too far from $\mu=M_X$, the 
relevant scale of the problem. 
The dashed (blue) curve essentially reproduces the results of 
Ref.~\cite{guo}. The solid (red) curve is our result, with the loop integrals 
regularized within the PDS scheme. The (gray) horizontal band is the LHCb 
experimental value~\cite{Aaij:2014ala}, with uncertainties added in 
quadrature. 
Though these results are quite different from each other and not yet properly 
renormalized, one is still able to refute the conclusions from 
Ref.~\cite{swanson,Aaij:2014ala}, that a ratio $R$ much larger than 
$\sim 10^{-3}$ would disfavor a molecular nature of $X$. 
Regarding the behavior of these two different curves, 
a few comments are in order. 
First, since $R$ is an observable, it should not depend on $\mu$. Therefore, 
Fig.~\ref{fig:Ratios} indicates the need of proper renormalization. Second, 
we draw attention to the logarithmic scale of the figure. In the PDS scheme 
adopted here, the dependence of $R^{\rm loop}$ on the renormalization scale 
$\mu$ is much stronger than in~\cite{guo}, which uses the standard 
$\overline{\rm MS}$-scheme. This is somehow expected, since PDS-regulated 
loops take into account power divergences that are set to zero in 
$\overline{\rm MS}$. Power divergences lead to a richer structure in the 
renormalization-group (RG) evolution path, allowing the existence of a 
non-trivial fixed point~\cite{Birse:1998dk} that describes the 
non-perturbative 
aspects of weakly-bound systems. Such RG constraints can induce a larger 
dependence of the short-distance contact couplings $C_{\psi}$'s on $\mu$, 
as shown in the following. Most of power divergences come from magnetic 
interactions, c.f. Eqs.~(24)--(28) of Ref.~\cite{guo}, meaning that they 
are more sensitive to short-distance physics. 
Note that there are a few examples in weakly-bound nuclear systems where the 
short-distance sensitivity of magnetic interactions is also 
observed---see, for instance, Refs.~\cite{Fernando:2011ts,Fernando:2015jyd}. 

The previous discussion points to the need of including explicitly the contact 
interactions from diagram $(f)$ and perform a proper RG analysis. 
For practical reasons, we find more convenient to impose the RG-constraint 
on the decay width, that is, 
\begin{equation}
\frac{\partial\Gamma}{\partial\mu}=0\,.
\label{eq:RG-cond}
\end{equation}
This condition is imposed, numerically, on each decay channel 
$\gamma J/ \psi$ and $\gamma \psi'$, tied to the experimental constraint 
$R\approx 2.46$~\cite{Aaij:2014ala}. 
Since both $\Gamma_{\gamma\psi'}$ and $\Gamma_{\gamma J/\psi}$, 
contrary to their ratio, are not well-determined, we choose a few 
representative values of $\Gamma_{\gamma J/\psi}$ as initial boundary 
condition in our RG-equation~(\ref{eq:RG-cond}). 
One finds two sets of solutions for the $\mu$-dependent contact terms, 
$C_{J/\psi}$ and $C_{\psi'}$, which reflects the fact that we impose the 
RG-constraint essentially on the modulus-squared of the decay amplitude. 
We present only one of these sets, since the other leads to decay widths 
of the order of tenths of MeV while the total decay width of the $X$ has an 
upper limit of $1.2$~MeV~\cite{Agashe:2014kda}. 
We assume that all $\mu$-dependence is given by the couplings 
$C_{\psi}$'s, ignoring eventual $\mu$-dependences on the couplings 
$x_{\rm nr}$ and $g_{\psi}$. 

\begin{figure}[hbt]
\centering
  \subfigure[fig-a][]{\includegraphics[width=7.5cm]{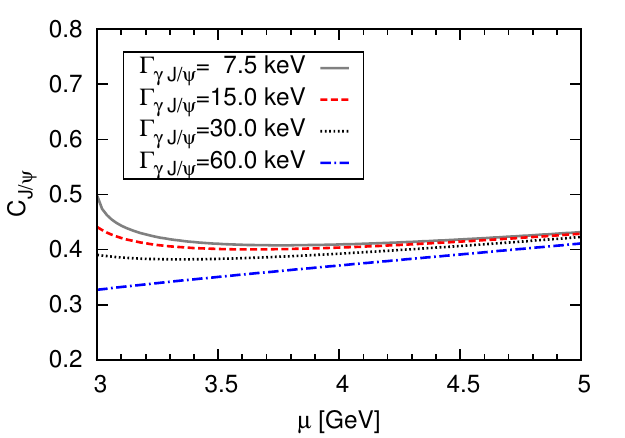}}
 \qquad
 \subfigure[fig-b][]{\includegraphics[width=7.5cm]{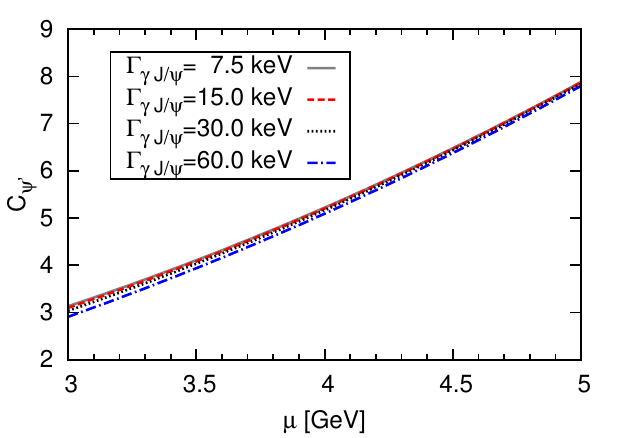}}
\caption{Short-distance contact interactions $C_{J/\psi}$ and $C_{\psi'}$, 
as functions of the renormalization scale $\mu$.}
\label{fig:Cterms}
\end{figure}
In figure \ref{fig:Cterms} we present our RG results for the 
short-distance couplings $C_{j \psi}$ and $C_{\psi'}$. 
As indicated, the curves correspond to different values of 
$\Gamma_{\gamma J/\psi}$. 
Within the selected ranges of $\mu$ and $\Gamma_{\gamma J/\psi}$, one 
notices a smooth variation on $C_{J/\psi}$, within $\approx 30$\%, about 
the same relative error on $R$ reported by LHCb. We stress that the 
extraction of the couplings $C_{\psi}$'s depend on $x_{\rm nr}$ via 
Eq.~(\ref{eq:Mloop}) and has a much larger relative theoretical uncertainty. 
On the other hand, $C_{\psi'}$ exhibits a stronger variation with $\mu$, 
meaning that $\Gamma_{\gamma\psi'}$ is more sensitive to the short-distance 
physics not dynamically taken into account by our effective theory. 
It is important to emphasize that, from the EFT point of view, short-distance 
physics means not only compact configurations like charmonium or tetraquark, 
but also heavier molecular states that are integrated-out from the effective 
theory, for instance, a virtual $D_s\bar D_s^{\star}$ contribution. 

\begin{figure}[hbt]
\centering
  \subfigure[fig-a][]{\includegraphics[width=7.5cm]{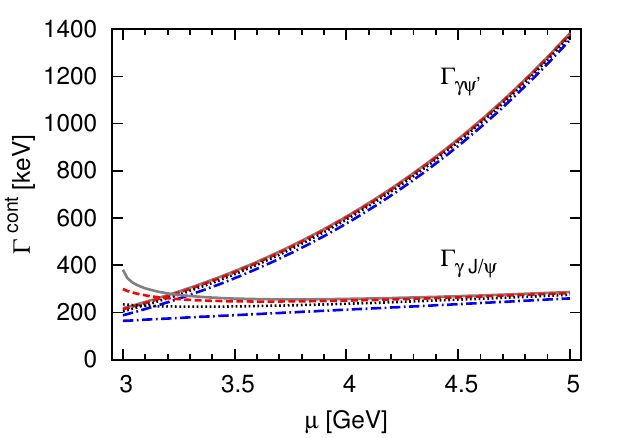}}
 \qquad
 \subfigure[fig-b][]{\includegraphics[width=7.5cm]{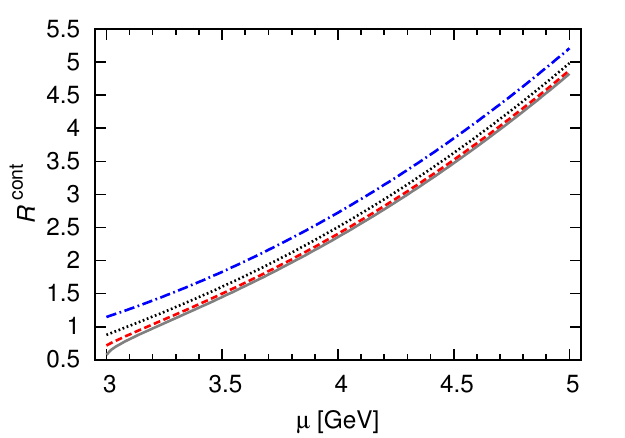}}
\caption{Contributions of the short-range interactions to the decay widths 
(a) and the ratio $R$ (b), as functions of the renormalization scale $\mu$. 
The curves follow the same labels from Fig.~\ref{fig:Cterms}.}
\label{fig:Gterms}
\end{figure}
To estimate the relevance of the short-distance interactions to $X$ radiative 
decays, we show in Fig.~\ref{fig:Gterms} only their contributions to the 
decay widths $\Gamma_{\gamma J/\psi}$ and $\Gamma_{\gamma\psi'}$, as well 
as the ratio $R$. 
To qualitatively interpret these results, let us focus on the dash-dotted 
(blue) curve of $\Gamma_{\gamma J/\psi}$. This has an input value of 60 keV. 
However, Fig.~\ref{fig:Gterms}(a) shows that its short-range interactions 
contribute about 200 keV. That implies a large cancellation between the 
long-(${\cal M}^{\rm loop}$) and short-(${\cal M}^{\rm cont}$) distance 
terms to generate the smaller width of 60 keV. In other words, the 
interplay between ${\cal M}^{\rm loop}$ and ${\cal M}^{\rm cont}$ is of 
a delicate balance, both having the same importance to the decay. 
This is even more dramatic in the case of $\Gamma_{\gamma\psi'}$. 
Such large cancellations between long- and 
short-distance terms may be a consequence of an underlying symmetry and is 
a question worth pursuing. 
We also checked that this qualitative cancellation holds for the 
$\overline{\rm MS}$-regulated loops, as much as for $\Gamma_{\gamma J/\psi}$ 
but less dramatic for $\Gamma_{\gamma\psi'}$. 
The $\mu$-dependence shown in figure \ref{fig:Gterms} may also be relevant 
in guiding theoretical models for the short-distance part.

In EFT approach, representations of short-range physics as contact 
interactions mean that all dynamical effects not explicitly taken into account 
may be relevant only beyond the EFT scale. That includes the opening of high 
momenta thresholds, leading to imaginary terms in the amplitude. Therefore, 
the constraint that the short-range coulplings $C_{\psi}$ remain real-valued 
within the energy range of the effective theory assures that there are no 
opening of high-energy thresholds. 
However, one notices numerically the appearance of an imaginary part on 
$C_{\psi}$ in the evolution of our RG-equation~(\ref{eq:RG-cond}), depending 
on the initial conditions. Choosing the range 
$3~{\rm GeV}\lesssim\mu\lesssim 5~{\rm GeV}$, to keep $C_{\psi}$ real one 
finds the restrictions 
\begin{equation}
\Gamma_{\gamma J/\psi}\gtrsim 7.5~{\rm keV}
\quad\mbox{and}\quad
\Gamma_{\gamma\psi'}\gtrsim 18.5~{\rm keV}\,. 
\label{eq:minwidth}
\end{equation}
Though the precise values depend on the EFT range, these numbers can 
be put to experimental scrutinity. 

\acknowledgments
The authors thank Kanchan Khemchandani and Alberto Martinez-Torres for 
discussions. This work was supported by CNPq and FAPESP. 

\bibliographystyle{apsrev4-1}

\end{document}